# Designing Artificial Two-Dimensional Landscapes *via* Room-Temperature Atomic-Layer Substitution

Yunfan Guo[1,‡,*], Yuxuan Lin[1,‡], Kaichen Xie[2,‡], Biao Yuan[3], Jiadi Zhu[1], Pin-Chun Shen[1], Ang-Yu Lu[1], Cong Su[4], Enzheng Shi[5], Kunyan Zhang[6], Zhengyang Cai[1], Jihoon Park[1], Qingqing Ji[1], Jiangtao Wang[1], Xiaochuan Dai[1], Xuezeng Tian[7], Shengxi Huang[6], Letian Dou[5], Ju Li[4], Yi Yu[3], Juan-Carlos Idrobo[7], Ting Cao[2], Tomás Palacios[1], Jing Kong[1,*]

[1]Department of Electrical Engineering and Computer Science, Massachusetts Institute of Technology, Cambridge MA 02139, United States

[2]Department of Materials Science & Engineering, University of Washington, Seattle, WA, 98195, United States

[3]School of Physical Science and Technology, ShanghaiTech University, Shanghai 201210, China.

[4]Department of Nuclear and Materials Science and Engineering, Massachusetts Institute of Technology, Cambridge MA 02139, United States

[5]Davidson School of Chemical Engineering, Purdue University, West Lafayette, Indiana 47907, United States.

[6]Department of Electrical Engineering, The Pennsylvania State University, University Park, Pennsylvania 16802, United States

[7]Center for Nanophase Materials Sciences, Oak Ridge National Laboratory, Oak Ridge, TN, 37831, United States

*E-mail: jingkong@mit.edu (J.K.); yfguo112@mit.edu. (Y.G.)

‡These authors contributed equally to this work.

**Manipulating materials with atomic-scale precision is essential for the development of next-generation material design toolbox[1-3]. Tremendous efforts have been made to advance the compositional[4,5], structural[6,7] and spatial accuracy of material deposition and patterning, including molecular beam epitaxy[8], atomic-layer deposition/etching[9], electron-beam lithography (EBL)[10], *etc*. The family of two-dimensional (2D) materials provides an ideal platform to realize atomic-level material architectures[11-13]. The wide and rich physics of these materials have led to fabrication of heterostructures[14-18], superlattices[19-21], and twisted structures[22-24] with breakthrough discoveries and applications. Here, we report a novel atomic-scale material design tool that selectively breaks and forms chemical bonds of 2D materials at room temperature, called atomic-layer substitution (ALS), through which we can substitute the top layer chalcogen atoms within the three-atom-thick transition-metal dichalcogenides (TMDs) using arbitrary patterns. Flipping the layer via transfer allows us to perform the same procedure on the other side, yielding programmable in-plane multi-heterostructures with different out-of-plane crystal symmetry and electric polarization. First-principle calculations elucidate how the ALS process is overall exothermic in energy and only has a small reaction barrier, facilitating the reaction to occur at room temperature. Optical characterizations confirm the fidelity of this design approach, while transmission electron microscopy shows the direct evidence of Janus structure and suggests the atomic transition at the interface of designed heterostructure. Finally, transport and Kelvin probe measurements on MoXY (X, Y = S or Se; X and Y corresponding to the bottom and top layers) lateral multi-heterostructures reveal the surface potential and dipole orientation of each region, and the barrier height between them. Our approach for designing artificial 2D landscape down to a single layer of atoms can lead to unique electronic, photonic and mechanical properties previously not found in nature. This opens a new paradigm for future material design, enabling unprecedented structures and properties for unexplored territories.**

Thermal chemical vapor deposition (CVD) is one of the most essential strategies to synthesis two-dimensional materials, in which thermal energy plays an indispensable role in both thermodynamics and kinetics during growth. On the other hand, reducing the processing temperature is a continuing imperative to explore new growth mechanism and unique material structures that are impossible to obtain from traditional temperature-controlled reactions. In this work, we designed artificial landscapes in 2D materials by using atomic-layer substitution (ALS), in which chemical bonds in monolayer TMD were selectively broken up and reformed at room temperature, making possible designer electric-dipole profiles of an in-plane multi-heterostructures. Unlike previous substitutional alloying methods at high temperature (700 °C-900 °C)[25,26], material conversions could be realized without extra thermal energy inputs by using the ALS strategy. As shown in Figure 1a, the presence of hydrogen radicals[27] produced by the remote plasma could easily strip the chalcogenide atoms on the top layer in a gentle and chemical way. Meanwhile, the low-pressure system facilitates the supply of vaporized substitutions to take the place of missing atoms, resulting in an asymmetric Janus structure of MXY (M=Mo or W, X=S or Se, Y=Se or S) at room temperature. This is also in contrast with the previous strategy[28], which depends on additional heating and annealing process to form new chemical bonds and remove excess selenium atoms. Most significantly, our room-temperature method makes it possible to implement arbitrary patterns in a 2D plane through any mask definitions (here, we use lithographic patterns as an example), enabling locally programmed out-of-plane atomic structures (Fig. 1b, where EBL can define arbitrary patterns using a polymethyl methacrylate (PMMA) mask). ALS will convert the exposed areas to MSSe, allowing for the lateral heterostructure between $MS_2$ and Janus MSSe within the same host monolayer (Fig. 1c). Furthermore, the chalcogen on the other side can be replaced by using a flip-over transfer and another selective-ALS conversion, resulting in lateral multi-heterostructures of $MS_2$-MSSe-MSeS-$MSe_2$ (Fig. 1d) with tailored functionalities.

As a proof of concept, we used CVD-synthesized monolayer $MoS_2$ and performed ALS to selenize the top sulfur layer and later the bottom layer (Extended Data Fig. 1). From the optical microscopy (OM) images (Extended Data Fig. 1b-d), the optical contrast of the $MoS_2$, MoSSe and $MoSe_2$ triangular single crystals remained uniform and intact through ALS, suggesting the process is benign to the 2D lattice. Atomic force microscopy (AFM) confirmed this (Extended Data Fig. 1e-g), as their surfaces remained pristine and flat after each ALS step. In addition, the single/double-sided conversions can naturally be extended to continuous large-area films (Extended Data Fig. 1h-j). From a material synthesis perspective, $MoSe_2$ is obtained from a $MoS_2$ template at room temperature, without the need of high temperature conversions[29,30], and thus avoids subsequent lattice or substrate damage. This provides an alternative route to obtain Se-based TMDs at room temperature. More importantly, this could be used to synthesize materials and structures that cannot be directly produced.

To confirm the substitution of the chalcogen layer by the ALS, optical characterizations provide a facile method due to the sensitivity of both Raman and photoluminescence (PL) spectroscopy to the crystal structure. For TMDs, the $A_{1g}$ and $E_{2g}$ mode correspond to the out-of-plane and in-plane lattice vibrations, respectively. Therefore, changing the top layer of chalcogens produces different phonon frequencies for the same type of lattice vibrations. As shown in Fig. 2a, the $MoS_2$ $A_{1g}$ (404 cm$^{-1}$) and $E_{2g}$ (383 cm$^{-1}$) modes shift to 288 cm$^{-1}$ and 355 cm$^{-1}$, respectively, in Janus MoSSe, due to the change of atomic mass and the broken symmetry in the vertical direction[28]. Further ALS on the other sulfur layer to full selenization yields a sharp peak at ~239 cm$^{-1}$ and a broad one at ~ 284 cm$^{-1}$, which are consistent with the $A_{1g}$ and $E_{2g}$ modes in monolayer $MoSe_2$[31]. This evidence supported by the strong PL emission peaks shown in Fig. 2b, which shows that the PL shift from 1.85 eV (pristine $MoS_2$) to 1.72 eV (Janus MoSSe) to 1.60 eV ($MoSe_2$). The PL positions are all in agreement with previous literature[28,31]. In regards to the uniformity, Raman mappings (Extended Data Fig. 2a-c) and PL intensity mappings (the insets of Fig. 2b) also confirm that the ALS has high spatial homogeneity.

With respect to many physical properties, the TMD $WX_2$ has attracted more attention due to W's stronger spin-orbit coupling and overall cleaner optical signatures. Compared with MoSSe, the larger elastic modulus and smaller carrier effective mass of WSeS/WSSe may give rise to higher carrier mobility, especially for holes[32,33]. Through the ALS process, we accomplished the conversion from $WS_2$ to Janus WSSe and $WSe_2$ (Extended Data Fig. 1k-m and Fig. 2g-i), with the corresponding PL emissions shift from 2.01 eV ($WS_2$) to 1.83 eV (WSSe) and to 1.63 eV ($WSe_2$) (Fig. 2d). In addition, to show its universality, we sulfurized monolayer $MSe_2$ (M=Mo and W) to Janus MSeS and to $MSe_2$ (Fig. 2c and Extended Data Fig. 2d-f). The evolutions of Raman peaks are all in consistent with the literature[33], which suggest the successful conversion by using ALS method.

With the ALS strategy, complex patterns which may be necessary for intricate electronics were generated in monolayer $MoS_2$. Since this kind of material conversion is performed in room temperature, a variety of materials can be used as masks to design patterns, such as polymers, nanowires, carbon nanobutes and *etc*. Here, as a proof of concept, we use EBL to define PMMA patterns such as the "MIT" logo and "Tim the beaver", the MIT mascot, on a continuous monolayer CVD-$MoS_2$ film. ALS converted the exposed regions to Janus MoSSe, while the masked regions remained as pristine $MoS_2$. Following the same aforementioned characterization scheme, OM (Fig. 2e), scanning electron microscopy (SEM) images (Fig. 2f), Raman and PL mappings (Fig. 2g, Extended Data Fig. 3a, b) all clearly verify that $MoS_2$ and MoSSe are well-located as designed. In principle, the minimum feature size of ALS method should be determined by the resolution of lithography (Extended Data Fig. 3c-i), with potential for scaling down using extreme ultraviolet (EUV) or helium-beam based lithography. Inverting the $MoS_2$/MoSSe heterostructures and repeating selective ALS on the other side yielded lateral multi-heterostructures with $MoS_2$, Janus MoSSe and MoSeS, and $MoSe_2$ (OM image in Extended Data Fig. 4a, AFM images in Extended Data Fig. 4b, c, Raman spectroscopy in Extended Data Fig. 4d., Raman mappings in Fig. 2h and Extended Data Fig. 4e-g). The ability to create uniform Janus and artificial TMDs multi-heterostructures within a continuous atomic-layer has never been achieved previously.

In order to verify the atomic structure of Janus MoSSe, we used annular dark-field scanning transmission electron microscopy (ADF-STEM) to obtain the tilted images of MoSSe. The image was obtained by rotating the MoSSe sample 10° along the direction given by the Mo-S and Mo-Se bonds, such that the Mo, Se and S atoms can be individually identified. As shown in Figure 2i, the tilted ADF-STEM image enables to separate the Se and S atoms. We find that the Se atoms are located on one side of the monolayer MoSSe, and S atoms on the opposite side, which is a direct evidence of the Janus structure. The corresponding intensity profile in Figure 2j clearly show the individual Mo, Se and S atoms with their total peak intensities proportional to their atomic numbers. As these Janus heterostructures are not grown from the edge of an existing crystal, characterizing the junction between these lateral heterostructures is highly critical to determining the fidelity or crystal quality of the resultant heterostructures. ADF-STEM shows that the $MoS_2$/MoSSe interface is seamlessly connected at the boundary defined by lithography (Fig. 2k and Extended Data Fig. 5a). As ADF-STEM intensity scales with the atomic number, the $MoS_2$ region with weaker intensity from S-S atomic pairs on the left side of Fig. 2k is clearly distinguished from the stronger intensity from S-Se atomic pairs of MoSSe on the right. Additionally, the different structures of $MoS_2$ and MoSSe are further verified by the intensity profile of electron scattering, where the atomic positions of Mo, S/Se, and 2S are corresponding to their gradually decreased intensities (Fig. 2i, m). The energy-dispersive X-ray spectroscopy (EDX) results (Extended Data Fig. 5c-h) indicate that Se and S atoms are well- confined in the selective regions. As the selected area electron diffraction (SAED) patterns show non-separable patterns from these two lattices

(Extended Data Fig. 6 a-d), real space strain mapping were further performed (Extended data Fig. 6g, h), in which the slight lattice difference in the MoSSe region can be observed compared to the reference $MoS_2$ regions.

To achieve a better understanding of the conversion process, we investigated the evolution of the Raman peaks during the synthesis. The results indicated in Fig. 3a, b and Extended Data Fig. 7a show that within 10 seconds, the $A_{1g}$ peak of Janus MoSSe begins to emerge (at a frequency ~10 cm$^{-1}$ lower than the final point, most likely due to the modulation of phonon frequency from the local phonon vibrations at the reaction onset[32]). The intensity of the Janus MoSSe $A_{1g}$ mode increases with longer conversion time; meanwhile the intensity of the $MoS_2$ $A_{1g}$ peak decreases and eventually vanishes in 15 minutes. For material conversion through high temperature substitution, previous reports[30] showed that the atomic replacement begins at energetically favorable sites (crystal edges, defect sites, grain boundaries) and gradually progresses to the other regions of the flakes, indicating it is controlled by thermodynamics. In contrast, the room temperature operation of ALS suggests it is a kinetic process, in which the energy needed for the atomic substitution is supplied by the highly reactive hydrogen radicals. This leads to the spatially homogeneous substitution of atoms over the whole flake since the reaction onset (Insets of Fig. 3b and Extended Data Fig. 7b), and the fast progression of the reaction with a time scale as short as 30 seconds (Fig. 3a and Extended Data Fig. 7a).

In addition to the time-dependent Raman studies, we aimed to understand the microscopic mechanism of the ALS, especially the roles of hydrogen plasma using *ab initio* calculations. Density functional theory (DFT), within the generalized gradient approximation, including van der Waals corrections (PBE-D2) was used on a supercell containing four $MoS_2$ unit cells (larger supercells are included in the SI). Based on the experimental conditions, we tested two reaction paths (Fig. 3c, d and Extended Fig. 8d), the conventional high temperature substitution (replacing S by Se by thermal activation), and the ALS process. Our calculations show that the remote plasma-assisted ALS has the lowest energy barrier and is overall exothermic due to the active roles of H radicals. Figure 3c exhibits that H radicals firstly adsorb on the upper surface of monolayer $MoS_2$ with adsorption energy of ~0.5 eV/H (which further increase with the H coverage as shown in Extended Data Fig. 8a). As the H coverage increases, two H atoms form bonds with one S atom (Extended Data Fig. 8b), and the $H_2S$ desorbs, leaving a S vacancy. Surprisingly, this desorption process only involves a small activation barrier < 0.5 eV (i.e., 12 kcal/mol), an order of magnitude smaller than the total bond energies of 3 Mo-S bonds (~ 6 eV). We note that in this reaction, the removal of S is unlikely caused by collisions with hydrogen radicals, because the kinetic energy of a H radical in the remote hydrogen plasma is about two orders of magnitude smaller than the displacement energy of a chalcogen atom[33] and the masses are very different between H and S atoms. The S vacancy is then occupied by a nearby Se atom. In contrast, the high-temperature pathway (Extended Data Fig. 8d) may lead to Se directly substituting S, involving an energy barrier of ~ 2.5 eV that occurs when the Mo-S bond breaks (Fig. 3e). This calculated energy barrier matches well with the experimental findings, where the high-temperature substitution happens at ~ 1000 K[26].In contrast, , the ALS approach reduces the reaction energy barrier that makes possible the room-temperature conversion, in which the reaction kinetics is controlled by the active H radicals.

A major application of 2D materials and their heterostructures lies in novel electronic or optoelectronic devices. With ALS, we found that high-quality materials are obtained, and the energy misalignment of band edges exist at both the MoSSe/MoS$_2$ and MoSe$_2$/MoSeS lateral heterojunctions with a type-II band alignment. Multiple back-gate transistors with long and short electrodes were fabricated and 4-probe measurements were used to eliminate the contact resistance (OM images in the insets of Fig. 4b, c). Typical transfer characteristics (channel current $I_D$ versus back-gate voltage $V_G$) are plotted in Fig. 4 a, and used to extract the threshold voltage $V_T$, defined as the intersection of the linear fit of the on-current and the x-axis, and the field effect mobility $\mu_{FE}$, defined as $\mu_{FE} = (L/W)g_m/(V_D C_{ox})$, where $L$ and $W$ are the channel length and channel width; $g_m$ is the transconductance, $V_D$ is the source-drain voltage (bias across the channel), and $C_{ox}$ = 12.1 nF/cm$^2$ (capacitance of the 285 nm SiO$_2$ gate oxide). The $V_T$ and $\mu_{FE}$ of tens of devices were measured for the starting CVD MoS$_2$, single-side-converted Janus MoSSe, as well as double-side-converted MoSe$_2$, summarized in the inset of Fig. 4a. Average $\mu_{FE}$ for MoS$_2$, Janus MoSSe, and ALS MoSe$_2$ are 8.94 cm$^2$/Vs, 2.11 cm$^2$/Vs and 0.17 cm$^2$/Vs, respectively. Average $V_T$ for them are -1.9 V, 36.0 V and 51.6 V, respectively. The gradual degradation of mobility from MoS$_2$ to MoSe$_2$ may arise from the introducing of defects through the ALS or transfer processes. Note that the mobility value measured in our Janus MoSSe devices is two orders of magnitude higher than previously reported values[34], which is indicative of the high quality of materials from this approach. Furthermore, the increasing of $V_T$ values indicates that MoS$_2$ is the most *n*-type doped, whereas

Janus MoSSe and MoSe$_2$ are decreasingly *n*-doped, in agreement with DFT calculated band alignment (Fig. 4f). This band alignment configuration further determines the electrical polarity of Janus MoSSe/MoS$_2$ and MoSe$_2$/Janus MoSeS lateral heterojunctions. The Janus MoSSe and the MoSe$_2$ regions are the anodes of these two n-n$^+$ junctions, respectively, and confirmed with output characteristics ($I_D$ versus $V_M$ with different $V_G$, where $V_M$ is the voltage drop across the two inner short electrodes (M1 and M2) as $I_D$ is applied across the two long electrodes (D and S)) as shown in Fig. 4b, c. The weak rectification behaviors observed on both lateral junctions suggest that the barrier heights (denoted as $\Phi_B$) at the lateral heterojunctions are very low. We performed temperature-dependent 4-probe transport measurement and used the thermionic emission model to extract the gate-dependent barrier heights (see Methods for details, and Extended Data Fig. 9 and Fig. 10), plotted in Fig. 4d, for $\Phi_B$=30 meV and 50 meV for MoSSe-MoS$_2$ and MoSe$_2$-MoSeS lateral heterojunctions, respectively, when the channels are turned on completely ($V_G$ > 50 V).

More uniquely, our ALS approach brings about the first programmable control of the out-of-plane electrical dipole in TMD materials (induced by the electronegativity difference between S and Se atoms in Janus MoSSe). We performed Kelvin probe force microscope (KPFM) measurements on the MoS$_2$-MoSSe-MoSeS-MoSe$_2$ grid samples (Fig. 4e), to observe clear surface potential differences among these four distinct materials. As expected, the Janus MoSSe region with the Se layers on top and the flipped MoSeS region with the S layers on top display the highest and the lowest surface potentials, respectively, whereas the surface potentials for the MoS$_2$ and MoSe$_2$ regions are in between. The measured potential difference between the top Se layer of the MoSSe region and the top S layer of the MoSeS region is around 100 meV. Such potential difference is even larger than that between the MoSe$_2$ and MoS$_2$ regions (which is around 15 meV), suggesting that the Janus MoSSe (or MoSeS) holds a large intrinsic dipole which is locked to the 2D surface normal. Note that this dipole can be selectively patterned to be zero (MoS$_2$, MoSe$_2$), positive (MoSSe), and negative (MoSeS) on such a 2D material "canvas", which would enable many novel nanostructures and devices with intriguing electrical and optoelectronic properties. For example, it is possible to make in-plane quantum wells, superlattices, or photonic devices by generating periodic arrays of such dipole/non-dipole lateral heterostructures with 1-100 nm periodicity. Not only can the transport and optical properties of this MoXY multi-heterostrutures be altered, any materials in close vicinity would also be modulated by such an electrostatic "canvas". Unusual physical properties and unprecedented functionality may be possible on such a platform, ranging from nonlinear optics, electronic band engineering, nanoscale origami, to electrochemical catalysis, all of which can be modulated by different electrostatic forces.

In summary, the ALS presented here is a powerful yet universal strategy to program material properties at the atomic-layer limit. Due to operating at room temperature, it allows for patterning which can create diverse artificial low-symmetry 2D materials and their heterostructures. ALS generates a seamless, high quality interface between different structures, and can complement existing vdW heterostructure fabrication techniques by adding a completely new class of materials (Janus and lateral Janus heterostructures). Furthermore, by designing the ALS-generated heterostructure pattern and surface charge distribution, great potential is anticipated in new physical discoveries and future applications.

**Extended Data** is available in the online version of the paper.

**Acknowledgements:** The preliminary experiments of this work are supported by the Air Force Office of Scientific Research under the MURI-FATE program, Grant No. FA9550-15-1-0514. The characterization of the Janus Materials at a later stage were supported by the U.S. Department of Energy (DOE), Office of Science, Basic Energy Sciences (BES) under Award DE-SC0020042. Y.L. and T.P. acknowledge the U.S. Army Research Office through the Institute for Soldier Nanotechnologies, under Cooperative Agreement number W911NF-18-2-0048, and the STC Center for Integrated Quantum Materials, NSF grant no. DMR 1231319. P.S. and A.L. acknowledge the funding from the Center for Energy Efficient Electronics Science (NSF Award No. 0939514) and the U. S. Army Research Office through the Institute for Soldier Nanotechnologies at MIT, under Cooperative Agreement No. W911NF-18-2-0048. K.X. and T.C. are partially supported by NSF through the University of Washington Materials Research Science and Engineering Center DMR-1719797. B.Y. and Y.Y. acknowledge the funding from NSFC (Grant No. 21805184), NSF Shanghai (Grant No. 18ZR1425200) and the Center for High-resolution Electron Microscopy (ChEM) at ShanghaiTech University (Grant No. EM02161943). C.S. and J. W. acknowledge the support through U.S. Army Research Office (ARO) under grant no. W911NF-18-1-0431. Q. J. acknowledge the support from the STC Center for Integrated Quantum Materials, NSF grant no. DMR 1231319. S.H. and K.Z. acknowledge the financial support from NSF (ECCS-1943895). L.D. acknowledges the support from U.S. Department of Defense, Office of Naval Research (Grant Number: N00014-19-1-2296). E.S. acknowledges the support from Davidson School of Chemical Engineering of Purdue University. J.L. and C.S. acknowledge the support by an Office of Naval Research MURI (Grant Number: N00014-17-1-2661). The crystallographic tilted STEM image research was conducted at the Center for Nanophase Materials Sciences, which is a DOE Office of Science User Facility (J.C.I.). We thank Y. Han, G. Cheng, N. Yao and N. Yan for helpful discussions.

**Author contributions:** Y.G. and J.K. designed the study; Y.G. synthesized and characterized the 2D artificial materials and heterostructures; Y.L. and P.S. performed device fabrication and measurements; T.C. and K. X. performed DFT simulation and interpreted the reaction mechanism; B.Y. and Y.Y. performed TEM characterization and data analysis; J.C.I. performed the crystallographic tilted STEM images and data analysis; C.S. and E.S. performed KPFM and SEM characterization and data analysis; J.Z., A.L., Z.C., J.P., Q.J., K.Z., J.W., X.D., X.T., S.H., L.D., J.L. and T.P. participated in data analysis and helpful discussion; Y.G., J.K., Y.L., T.C. wrote the manuscript; all authors read and revised the manuscript.

**Competing interests:** Authors declare no competing interests.

**Materials & Correspondence** should be addressed to J.K. and Y.G.

**Data and materials availability**: All data are available in the manuscript or supplementary information. All materials are available upon request to J.K.

## Methods

**Synthesis of MoS$_2$:** A target SiO$_2$ substrate was suspended between two other SiO$_2$/Si substrates with predeposited perylene-3, 4, 9, 10-tetracarboxylic acid tetrapotassium (PTAS) solution. All of these substrates were placed face-down on a crucible containing MoO$_3$ precursor in a 1-inch quartz tube. This crucible was placed in the middle of the heating zone with another sulfur crucible on the upstream. Before heating, the whole CVD system was purged with 1,000 sccm Ar (99.999% purity) for 3 min. Then, 20 sccm Ar was introduced into the system as a carrier gas. The growth system was heated to 625 °C for 15 min. The MoS$_2$ growth was carried out around 620 °C to 630 °C for 3 min under atmospheric pressure. After growth, the whole system was naturally cooled down to room temperature.

**Plasma-assisted atomic-layer-substitution (ALS):** As shown in Fig. 1a, we use a remote commercial inductively coupled plasma (ICP) system to substitute the top-layer sulfur atoms of monolayer MS$_2$ (M=Mo, W) with selenium. The CVD grown monolayer MoS$_2$/WS$_2$ were placed in the middle of a quartz tube. The plasma coil made by a cylindrical copper tube placed at the upstream of CVD furnace. Whenever masks were needed, the sample was covered with PMMA patterns defined by electron-beam lithography (EBL). At the beginning of the process, the whole system was pumped down to 5 mTorr to remove air in the chamber. Then, 5 sccm hydrogen was introduced into the system and the plasma generator was ignited with a power of 50 W. The hydrogen atoms assist the removal of the sulfur atoms on the top layer of $MoS_2$, at the same time, the vaporized selenium filled in the vacancy of the sulfur atoms, resulting in the asymmetric Janus structure of MoSSe. The whole process was performed at room temperature. After the reaction, the whole system was purged with Ar gas (99.999% purity) in 100 sccm to remove the residual reaction gas, and the pressure was recovered to atmospheric.

**Transfer:** For normal transfer, the samples were spin-coated with PMMA as a supporting layer. Then, they are put in the KOH solution and the PMMA/2D material was detached form the growth substrate and was floated on the surface of KOH solution. Afterwards, it was taken to a DI water bath by a glass slide and washed for several times, and then picked up with the target substrate. After baking on a hot plate at 100 °C for 15 minutes, the PMMA layer was removed with acetone and isopropanol (IPA).

**Raman and photoluminescence spectroscopy:** Raman and PL spectra were performed on a Horiba Jobin-Yvon HR800-confocal Raman Spectrometer. The laser excitation wavelength for Raman and PL measurements was 532.5 nm. A 100X objective was used to focus the laser beam. The laser power on the sample was about 0.1 mW. For PL and Raman mapping, the scanning step size is ~0.5 μm.

**Transmission electron microscopy:** The atomic structure of $MoS_2$-MoSSe lateral heterostructure was acquired using JEOL TEMs. The ADF-STEM image was taken on a 300 kV aberration-corrected JEOL GrandARM. The diffraction patterns were obtained on a 200 kV JEOL JEM-2100plus. The EDS mapping was collected on a 120 kV JEOL JEM-1400. The ADF-STEM experimental conditions were: condenser lens aperture 20 μm, probe size 9c, the probe current 7.3 pA, and camera length 15 cm, which corresponds to an inner collection angle of 42 mrad and outer collection angle of 180 mrad. The strain mapping is measured based on the geometric phase analysis using MacTempasX[35] software. The crystallographic tilted STEM images were obtained in an aberration-corrected Nion UltraSTEM 100™ operated at 100 kV, using a convergence semiangle of 30 mrad.

**KPFM:** KPFM measurements were performed using a Cypher AFM system (Asylum Research) with Ti/Ir-coated cantilevers (ASYELEC-01, Oxford Instruments) with a nominal mechanical resonant frequency and spring constant of 70 kHz and 2 N/m, respectively. The two-pass scanning method consists of a surface topography scanning and a nap mode which lifts up the tip and keep at a constant distance at 30 nm while scanning. During the nap mode, an AC bias is applied to the tip for tip vibration, and a DC voltage is applied between the tip and sample to minimize the vibration amplitude. The signal of DC voltage can be used qualitatively as a reflection of surface potential change, but there are still other factors may influence the tip vibration.

**Fabrication and transport measurements:** The as-grown MoSSe-MoS$_2$ or MoSe$_2$-MoSeS heterostructures were first transferred onto a 285 nm SiO$_2$/Si substrate. An EBL step and an electron-beam evaporation step (30 nm Ni/30 nm Au) were used to define the metal contacts. Another EBL step and a reactive ion etching process (O$_2$ plasma) were performed to etch the 2D material and define the channel regions. The transport measurements were carried out using a semiconductor parameter analyzer (Agilent 4155C) in a cryogenic probe station (Lakeshore) with a temperature controller and liquid nitrogen cooling. The pressure of the chamber was kept below 5×10$^{-6}$ torr.

To extract the barrier height at the lateral heterojunctions, a reverse-bias thermionic emission model is used[36]:

$$\ln\left(\frac{|I_R|}{T^{3/2}}\right) = \ln(A^*) - \frac{q\Phi_B}{k_B T}$$

Here $I_R$ is the reverse bias (-0.1 V) current, $T$ is the temperature, $A^*$ is the effective Richardson's constant, $k_B$ is the Boltzmann constant, $q$ is the electron charge, and $\Phi_B$ is the barrier height. The output characteristics of the MoSSe-MoS$_2$ and the MoSe$_2$-MoSeS lateral heterojunctions are plotted in Extended Data Fig. 9. The Arrhenius plot ($\ln(|I_R|/T^{3/2})$ versus 1000/$T$) is plotted in Extended Data Fig. 10.

For the 4-probe measurements, a drain-to-source voltage $V_D$ was applied to the device, and the drain current $I_D$ and the voltage drop $V_M$ across the two short electrodes (M1 and M2) was measured. Because there are no current following into M1 and M2, there is no voltage drop on the M1-channel and M2-channel contact ($V=IR$). As a result, the measured $V_M$ is completely from the channel part (from M1 to M2). In this way, the influence of the contact resistance can be eliminated if we plot $I_D$ versus $V_M$.

**Theoretical Calculations:** Ab initio calculations were performed using density functional theory in the Perdew–Burke–Ernzerhof exchange-correlation functional with dispersion correction (PBE-D2), implemented in the Quantum Espresso package[37]. A supercell arrangement was used with the cell dimension in the out-of-plane direction set at 20 Å to avoid interactions between the transition metal dichalcogenide layers and its periodic images. Both 2×2 and 3×3 supercells have been used. We use ultrasoft pseudopotentials with a plane-wave energy cutoff of 40 Ry. The structures were fully relaxed until the force on each atom is <0.005 eV/Å. Spin-orbit coupling was not included in our reaction path calculations.

In the band alignment calculations, we use norm-conserving pseudopotentials with a plane-wave energy cutoff of 80 Ry. We note that the Kohn–Sham band gaps from DFT neglect the quasiparticle self-energy corrections. However, the self-energies corrections to the band gaps in $MoS_2$ and $MoSe_2$ have been shown to be very similar[38]. As a result, the conduction (or valence) band alignment should be relatively insensitive to the self-energy corrections.

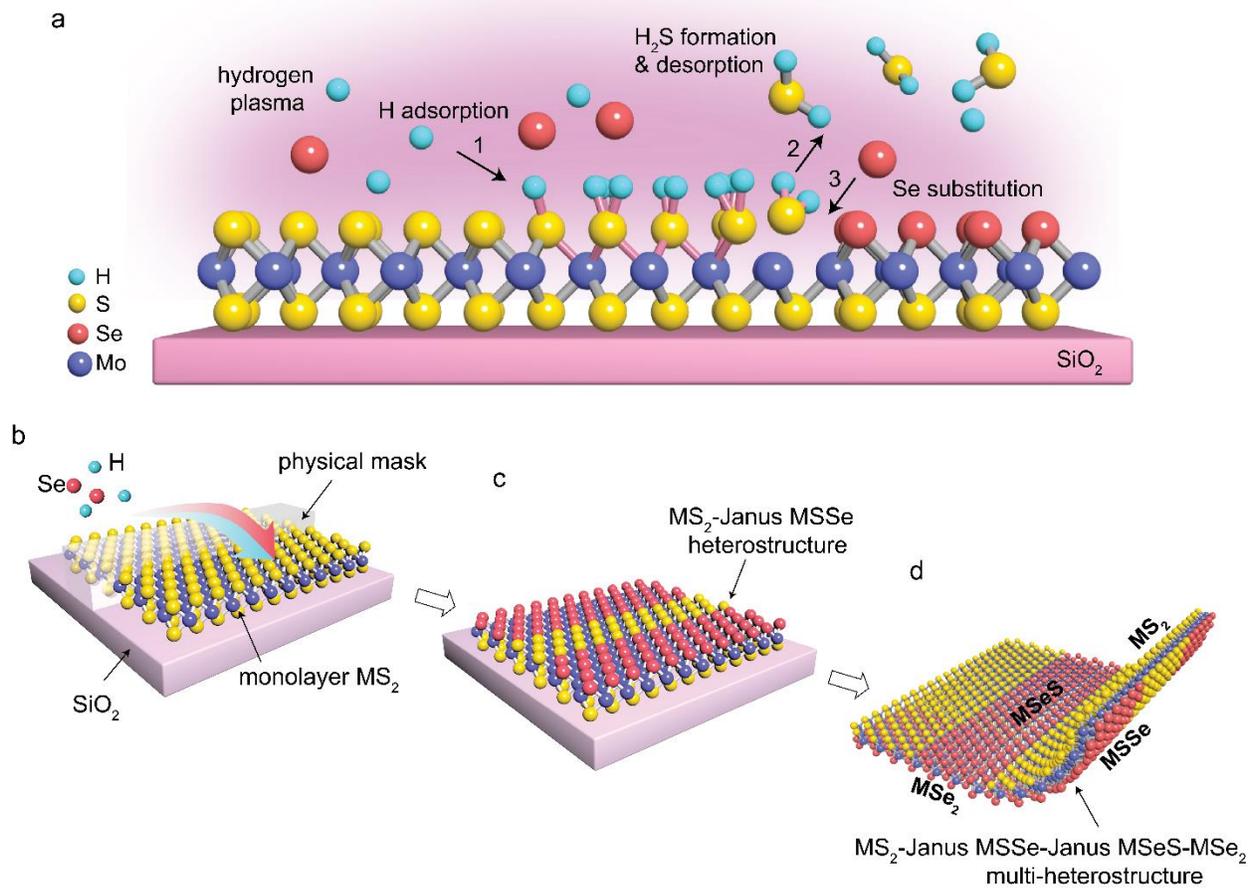

**Fig. 1| Room-temperature atomic-layer substitution (ALS) with programmable design in monolayer transition metal dichalcogenide. a,** Schematic illustration of the room temperature atomic-layer substitution from monolayer MoS$_2$. **b,** Schematic representation of programmable atomic-layer substitution process by lithography patterning and ALS process. **c,** Programmable monolayer MoS$_2$-Janus MoSSe heterostructure on SiO$_2$/Si substrate. **d,** A multi-heterostructure composed of monolayer MoS$_2$-Janus MoSSe-Janus MoSeS-MoSe$_2$ regions.

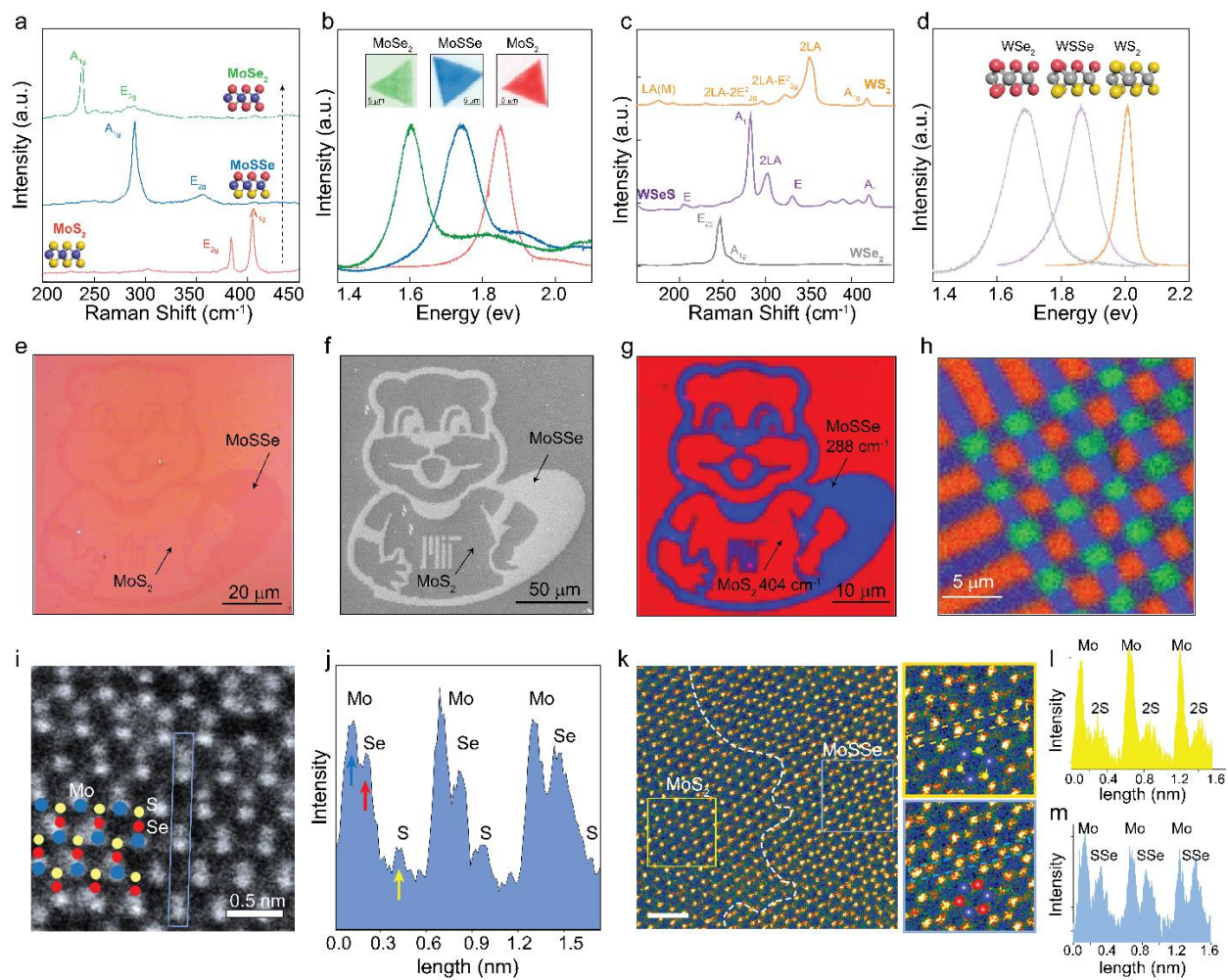

**Fig. 2| Characterizations of the atomic-layer substitution for monolayer TMD and their heterostructures. a,** Raman spectra of monolayer $MoS_2$, Janus MoSSe and $MoSe_2$. The monolayer Janus MoSSe and $MoSe_2$ were converted from $MoS_2$ by room-temperature ALS. Insets, crystal structures of $MoS_2$, Janus MoSSe and $MoSe_2$, respectively. **b,** Photoluminescence (PL) spectra of monolayer $MoS_2$, Janus MoSSe and $MoSe_2$. Insets, spatially resolved PL mappings of $MoS_2$, Janus MoSSe and $MoSe_2$ nanoplates at 1.85 eV, 1.72 eV and 1.60 eV, respectively. **c,** Raman spectra of monolayer $WSe_2$, Janus WSeS and $WS_2$. The monolayer Janus WSeS and $WS_2$ were converted from $WSe_2$ by ALS method. **d,** PL spectra of monolayer $WS_2$, Janus WSSe and $WSe_2$. Insets, the crystal structures of $WS_2$, Janus WSSe and $WSe_2$, respectively. **e,** The OM image of a Janus MoSSe-filled MIT mascot "Tim the Beaver" on a $MoS_2$ film pattern. There is obvious optical contrast between these two regions. **f,** SEM image of a MIT mascot "Tim the beaver" pattern (Janus MoSSe) on a $MoS_2$ canvas. **g,** Spatially resolved Raman mappings for $A_{1g}$ mode intensity of Janus MoSSe and $MoS_2$. **h,** Spatially resolved Raman mapping for $A_{1g}$ mode intensity of monolayer multi-heterostructures based on $MoS_2$-Janus MoSSe-Janus MoSeS-$MoSe_2$. The red area stands for MoS2 region, the blue one for Janus region and the green one for MoSe2. **i,** Tilted ADF-STEM image of a MoSSe sample to visualize the asymmetric atom structure in the vertical

direction. The corresponding Mo, Se and S atoms are schematically shown with blue, red and yellow circles, respectively. **j,** The intensity profile for the atomic chain highlighted in blue in (i) shows the intensity of individual Mo, Se and S atoms. **k,** Pseudocolor ADF-STEM image of a $MoS_2$-MoSSe interface (left panel), and the zoomed-in views of a $MoS_2$ region (right top panel, highlighted in yellow) and a Janus MoSSe region (right bottom panel, highlighted in blue), and their respective intensity profiles in (l) and (m).

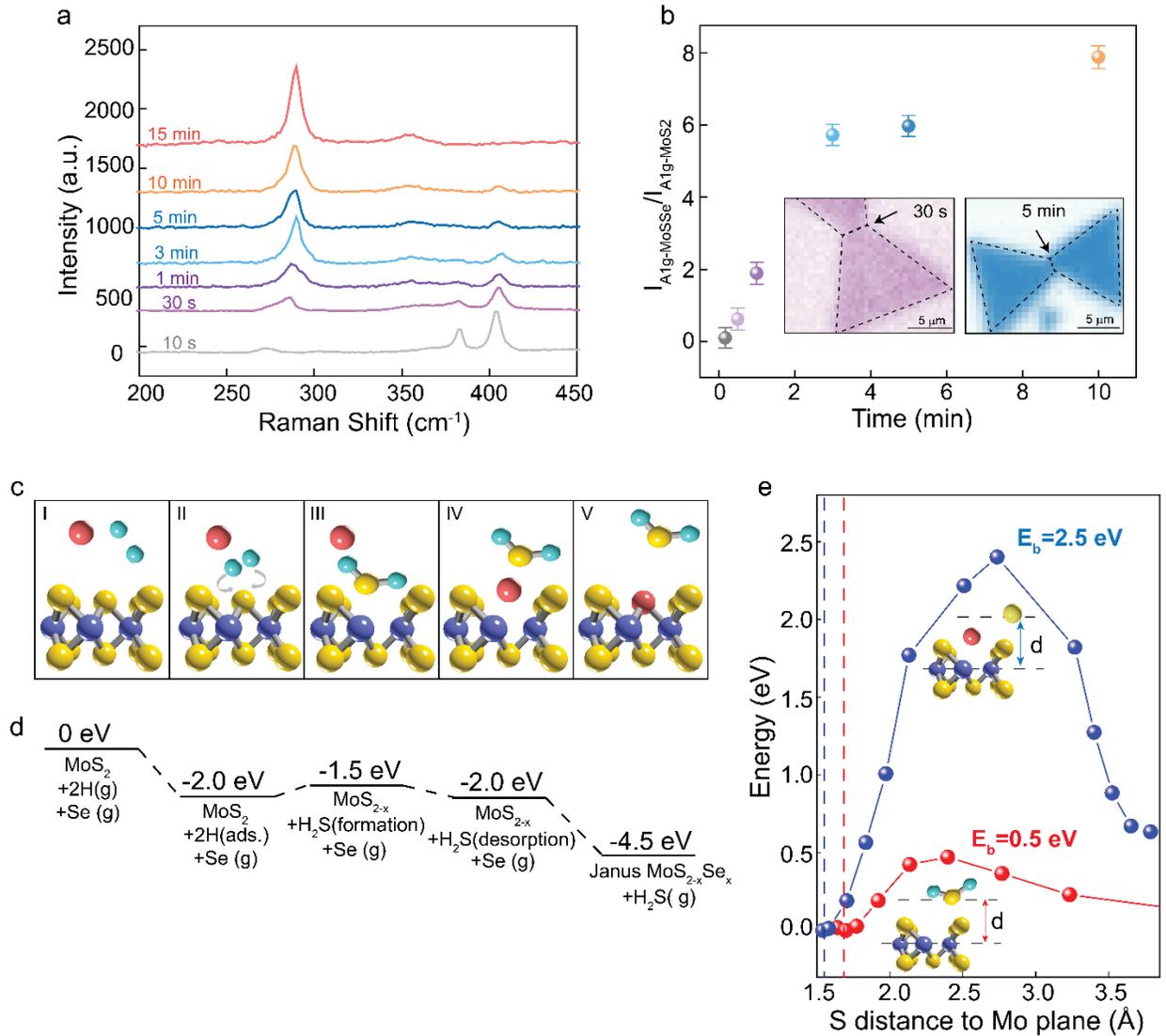

**Fig. 3| Mechanism study on the room-temperature conversion from monolayer MoS₂ to Janus MoSSe. a,** Raman spectra collected at several monolayer Janus MoSSe samples treated with hydrogen plasma for 10 s, 30 s, 1 min, 3 min, 5 min, 10 min and 15 min, respectively. **b,** The relationship between the conversion time and the corresponding intensity $I_{A1g\text{-}MoSSe}/I_{A1g\text{-}MoS2}$ ratio over the conversion process. Inset: Spatially-resolved Raman mappings of two coalesced Janus MoSSe flakes converted within 30 s, which show the uniform Raman intensity over the whole MoSSe regions. **c,** Schematics of the five key reaction steps for remote plasma-assisted ALS process (cartons from left to right): before H adsorption, 2 H adsorption and diffusion to the same S, formation of H₂S, desorption of H₂S, and Se occupation of the S vacancy. Purple, yellow, red, and green balls are Mo, S, Se, and H atoms, respectively. **d,** Free energy of each steps in (**c**), relative to that of the first step. **e,** Comparison of activation energy barrier between the non-thermal plasma-assisted atomic-layer substitution (red) and the conventional high-temperature substitution (blue). Dashed line indicates equilibrium S to Mo plane distances in each case. In the non-thermal

ALS process, the largest energy barrier occurs at a critical distance where H$_2$S starts to form. In high-temperature substitution, the largest energy barrier occurs when S and Se both disconnect with Mo.

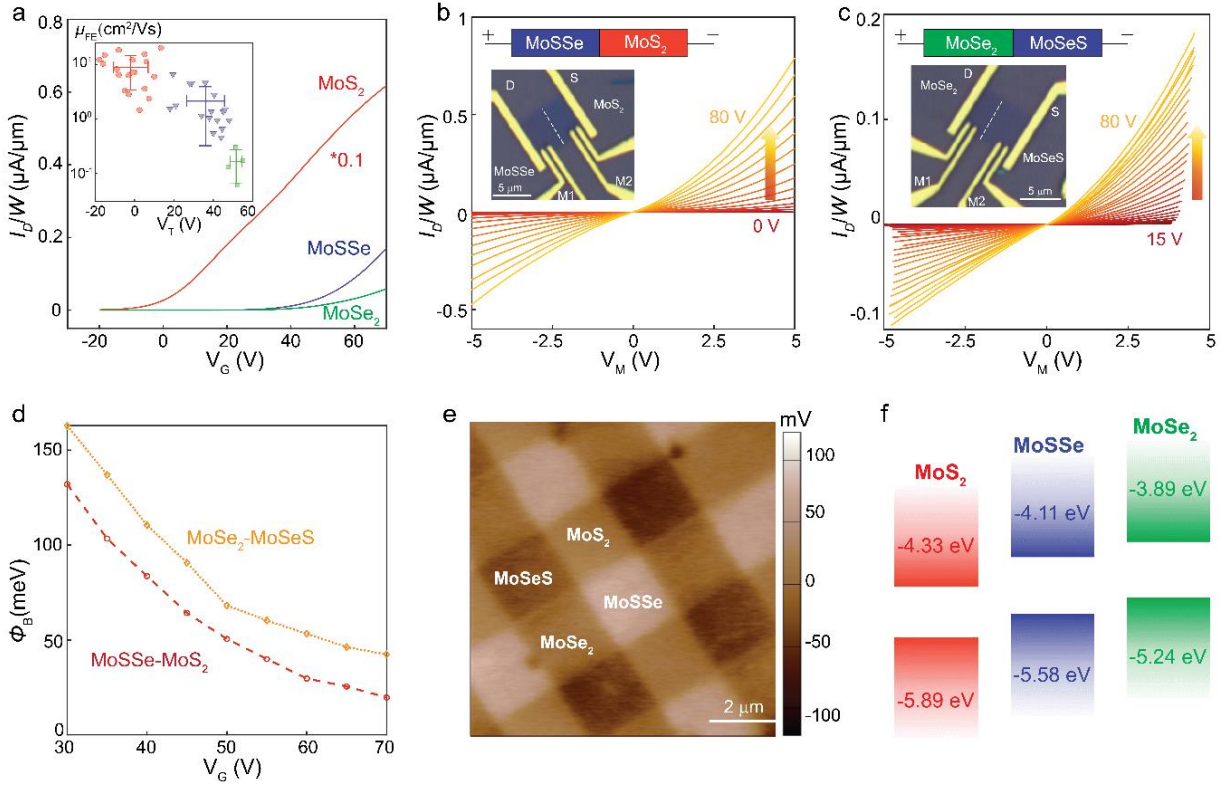

**Fig. 4| Electrical properties of MoS$_2$, Janus MoSSe, MoSe$_2$, and their lateral heteorstructures produced by ALS. a,** Transfer characteristics (current density $I_D/W$ versus back-gate voltage $V_G$, $V_D = 1$ V) of typical MoS$_2$, single-side-converted Janus MoSSe, and double-side-converted MoSe$_2$. Inset: field effect mobility $\mu_{FE}$ versus threshold voltage $V_T$ for tens of devices. The average mobility values for MoS$_2$, Janus MoSSe, and MoSe$_2$ are 8.94, 2.11 and 0.17 cm$^2$/Vs, and the average $V_T$ for them are -1.9, 36.0 and 51.6 V, respectively. **b-c,** 4-probe output characteristics (current density $I_D/W$ versus bias voltage $V_M$ with various back-gate voltage $V_G$) for (a) MoSSe-MoS$_2$ and (b) MoSeS-MoSe$_2$ lateral heterojunctions. The insets are the diagrams and OM images of these devices. **d,** Barrier height $\Phi_B$ as a function of $V_G$ for the MoSSe-MoS$_2$ (red) and the MoSeS-MoSe$_2$ (orange) lateral heterojunctions. **e,** KPFM image of the MoS$_2$-MoSSe-MoSeS-MoSe$_2$ multi-heterostructure. The Janus MoSSe region with the Se layer on top and the MoSeS region with the S layer on top display the highest and the lowest surface potential, respectively, whereas the surface potentials for the MoS$_2$ and MoSe$_2$ regions are in between. **f,** Conduction band minima and valence band maxima energies of MoS$_2$, Janus MoSSe, and MoSe$_2$ relative to vacuum, obtained from DFT calculations. For Janus MoSSe, the vacuum energy levels from upper and lower planes are averaged.